\documentclass[twocolumn,showpacs,aps,prl,superscriptaddress]{revtex4}
\usepackage{graphicx} 
\usepackage{dcolumn}
\usepackage{epsfig}
\usepackage{amsmath}

\newcommand{\BaBarPubYear}    {09}
\newcommand{\BaBarPubNumber}  {007}
\newcommand{\SLACPubNumber} {13600}
\newcommand{\LANLNumber} {0905.0868}

\newcommand{\BaBarType}      {PUB}  
\input babarsym   
\RequirePackage{xspace}

\newcommand{\pvec}{{\bf p}}

\newcommand{\calB}{\ensuremath{{\cal B}}}
\newcommand{\calT}{\ensuremath{{\cal T}}}

\newcommand{\bfemseven}{${\cal B}(10^{-7}$)}

\newcommand{\DE}{\ensuremath{\Delta E}}

\newcommand{\xf}{\ensuremath{{\cal F}}}

\newcommand{\thetaT}{\ensuremath{\theta_{\rm T}}}
\newcommand{\costhr}{\ensuremath{\cos\thetaT}}

\newcommand{\dEdx}{\ensuremath{\mathrm{d}E/\mathrm{d}x}}

\newcommand\etal{{\it et al.}}
\newcommand{\half}{\ensuremath{{1\over2}}}

\newcommand{\bma}[1]{\boldmath{$#1$}}

\newcommand{\bfig}{\begin{figure}[htbpc!]}
\newcommand{\efig}{\end{figure}}
\newcommand\bef{\begin{figure}}
\newcommand\edf{\end{figure}}
\newcommand\dbline{\noalign{\vskip 0.10truecm\hrule}\noalign{\vskip 2pt}\noalign{\hrule\vskip 0.10truecm}}
\providecommand{\tbline}{\noalign{\vskip 0.05truecm\hrule\vskip0.05truecm}}

\newcommand\beq{\begin{equation}}
\newcommand\eeq{\end{equation}}
\newcommand\bear{\begin{array}}
\newcommand\enar{\end{array}}
\newcommand\beqa{\begin{eqnarray}}
\newcommand\eeqa{\end{eqnarray}}
\newcommand\ben{\begin{enumerate}}
\newcommand\een{\end{enumerate}}

\newcommand{\UfourS}{\ensuremath{\Upsilon(4S)}}

\newcommand{\etagg}{\ensuremath{\eta_{\gaga}}}
\newcommand{\etappp}{\ensuremath{\eta_{3\pi}}}

\newcommand{\etapepp}{\ensuremath{\etapr_{\eta\pi\pi}}}

\newcommand{\etaprg}{\ensuremath{\etapr_{\rho\gamma}}}

   \newcommand{\rhoz}{\ensuremath{\rho^0}}

\newcommand{\fpizksks}{\ensuremath{\piz \KS \KS}}
\newcommand{\fetaksks}{\ensuremath{\eta \KS \KS}}
\newcommand{\fetaprksks}{\ensuremath{\etapr \KS \KS}}
\newcommand{\fetapretaprk}{\ensuremath{\etapr \etapr K}}
\newcommand{\fpizpizks}{\ensuremath{\piz \piz \KS}}
\newcommand{\fksksks}{\ensuremath{\KS \KS \KS}}

\newcommand{\fetaggksks}{\ensuremath{\etagg \KS \KS}}
\newcommand{\fetapppksks}{\ensuremath{\etappp \KS \KS}}

\newcommand{\fetapreppksks}{\ensuremath{\etapr_{\eta \pi \pi} \KS \KS}}
\newcommand{\fetaprrgksks}{\ensuremath{\etapr_{\rho \gamma} \KS \KS}}

\def\Lik         {{\ensuremath{\cal L}\xspace}}

\newcommand{\psiKs}{\mbox{$B^0\ra J/\psi  K^0_S $}}

\newcommand{\signf}{$\cal S$($\sigma$)}
\newcommand{\signff}{$\cal S$}
\newcommand{\eff}{$\epsilon$ (\%)}

\begin{document}

\begin{flushleft}
~\\
~\\
\end{flushleft}

\begin{flushright}
~\\
~\\
\babar-\BaBarType-\BaBarPubYear/\BaBarPubNumber \\
SLAC-\BaBarType-\SLACPubNumber \\
hep-ex/\LANLNumber
\end{flushright}

\title{
 \large \bf\boldmath Search for $B^0$ meson decays to \fpizksks,
 \fetaksks, and \fetaprksks 
}

%
\author{B.~Aubert}
\author{Y.~Karyotakis}
\author{J.~P.~Lees}
\author{V.~Poireau}
\author{E.~Prencipe}
\author{X.~Prudent}
\author{V.~Tisserand}
\affiliation{Laboratoire d'Annecy-le-Vieux de Physique des Particules (LAPP), Universit\'e de Savoie, CNRS/IN2P3,  F-74941 Annecy-Le-Vieux, France}
\author{J.~Garra~Tico}
\author{E.~Grauges}
\affiliation{Universitat de Barcelona, Facultat de Fisica, Departament ECM, E-08028 Barcelona, Spain }
\author{M.~Martinelli$^{ab}$}
\author{A.~Palano$^{ab}$ }
\author{M.~Pappagallo$^{ab}$ }
\affiliation{INFN Sezione di Bari$^{a}$; Dipartimento di Fisica, Universit\`a di Bari$^{b}$, I-70126 Bari, Italy }
% ...
\collaboration{The \babar\ Collaboration}
\noaffiliation

\date{\today}

\begin{abstract}
We describe searches for \Bz\  meson decays to the  charmless final
states \fpizksks, 
 \fetaksks, and \fetaprksks. 
 The data sample corresponds to 
$467 \times 10^{6}$  \BB\ pairs produced in 
\epem\ annihilation and collected with the
\babar\ detector at the SLAC National Accelerator Laboratory. 
We find no significant signals and determine the 90\% confidence
 level upper limits on the branching fractions, in units of $10^{-7}$,
$\calB(\Bz \to \fpizksks) <9 $,  $\calB(\Bz \to
\fetaksks) <10$, and 
 $\calB(\Bz \to \fetaprksks) <20$.
\end{abstract}

\maketitle

The observation of mixing-induced \CP\ violation
in \psiKs\ decays~\cite{MixInd}, as well as in the
charmless penguin-diagram 
dominated $\Bz \ra \etapr K^0$ decays~\cite{EtapKs}, and
of direct \CP\ violation both in the neutral
kaon system \cite{kaon} and in  $B^0\ra K^+\pi^-$
decays \cite{dir}, are in  agreement with predictions of the standard model 
(SM)  of electroweak interactions~\cite{SM}.  Further 
information  about  \CP violation and hadronic $B$ decays
can be provided by 
the measurement of branching fractions and time-dependent \CP\ asymmetries 
in $B$ decays to three-body final states containing two identical neutral 
spin zero particles and another \CP\ eigenstate spin zero 
particle~\cite{Tim}. \CP violating asymmetries have already been
measured in \Bz decays to \fksksks\ \cite{KsKsKs}  and to 
\fpizpizks\ \cite{pizpizks}, and a search has been performed in $B \to
\fetapretaprk$~\cite{EtapEtapK}.  
Other examples, in which study of time-dependent \CP\ violation 
asymmetry might be particularly interesting, 
are the \Bz\ decays to \fpizksks, \fetaksks, 
and  \fetaprksks. There are no theoretical estimations
for the branching fractions of  these SM-suppressed decay
modes. Contributions from physics beyond the SM may 
appear in these decays. 

Among $B$ meson decays to final states containing two kaons and an
additional light meson, only  $B^+ \ra K^+ K^- \pi^+$ has been
observed, with a branching fraction of $(5.0 \pm 0.5 \pm 0.5) \times
10^{-6}$~\cite{PreSoni}.  In this analysis an unexpected peak was 
 observed around $1.5$\gevcc in the $K^+K^-$ invariant-mass spectrum.
 Studies of decays with two neutral or charged kaons in the final
 state, such as those presented herein, may help to elucidate the
 nature of this structure~\cite{Soni}. 

We present the results of searches for neutral $B$ decays
 to charmless final states  \fpizksks,  \fetaksks\ and
 \fetaprksks, 
 which are studied for the first time.
The results are based on data collected
with the \babar\ detector~\cite{BABARNIM}
at the PEP-II asymmetric-energy $e^+e^-$ collider
located at the SLAC National Accelerator Laboratory. We use
an integrated luminosity of 426~\invfb, corresponding to 
$467 \times 10^6$ \BB\ pairs, recorded at the
$\Upsilon (4S)$ resonance (center-of-mass energy $\sqrt{s}=10.58\
\gev$) and, for the study of the background, 
44~\invfb collected 40~\mev\ below the resonance
(off-peak).

Charged particles from the \epem\ interactions are detected, and their
momenta measured, by a combination of five layers of double-sided
silicon microstrip detectors and a 40-layer drift chamber.
Both systems operate in the 1.5~T magnetic field of a superconducting
solenoid. Photons and electrons are identified with a CsI(Tl) crystal
electromagnetic calorimeter.
Charged particle
identification is provided by the average energy loss (\dEdx) in
the tracking devices and by an internally reflecting, ring-imaging
Cherenkov detector covering the central region (DIRC).
A $K/\pi$ separation of better than four standard deviations ($\sigma$)
is achieved for momenta below 3~\gevc.
Detector details may be found elsewhere~\cite{BABARNIM}.

The $B$ daughter candidates are reconstructed through their dominant decays:
$\eta\ra\gaga$ (\etagg), $\eta\ra \pi^+\pi^-\piz$ (\etappp) where
$\piz\to\gamma\gamma$, 
$\etapr\ra\eta\pip\pim$ (\etapepp) where
$\eta\ra\gaga$, and  
$\etapr\ra\rhoz\gamma$ (\etaprg) where $\rhoz\ra\pip\pim$. 
We require the laboratory energy of the photons to be  greater than 
30~\mev\ for \piz\ in \etappp, 50~\mev\ for \etagg\ in \etapepp,
and 100~\mev\  for \etaprg, and for 
\piz\ and $\etagg$ produced directly from the $B$ decay.
We impose the following requirements on the invariant mass (in \mevcc) of
the candidate final states:
$120 < m(\gamma\gamma) < 150$ for \piz,
$510 < m(\gamma\gamma) < 585$ for $\etagg$ produced directly from the $B$ decay,
$490 < m(\gamma\gamma) < 600$ for $\etagg$ in \etapepp,
$538 < m(\pip\pim\piz) < 558$ for \etappp,
$945 < m(\pip\pim\eta) <970$ for \etapepp,
$930 < m(\pip\pim\gamma) <980$ for \etaprg, and   $470 < m(\pip\pim)
<980$ for \rhoz.
Tracks from $\eta$ and \etapr\ candidate decays are rejected if
 their particle identification signatures from the DIRC and \dedx\ are
 consistent with those of protons, kaons, or electrons. 
Candidate \KS\ decays are formed from pairs of oppositely charged tracks 
with $486 < m(\pip\pim)<510$ \mevcc,  a decay vertex
$\chi^2$ probability larger than $0.001$,  and a reconstructed decay length 
greater than three times  its uncertainty.  

We reconstruct the $B$ meson candidate by combining two  
\KS\ candidates and a \piz, $\eta$,  or \etapr candidate. 
From the kinematics of the \UfourS\ decays we determine the
energy-substituted mass 
$\mes=\sqrt{\frac{1}{4}s-\pvec_B^2}$
and the energy difference $\DE = E_B-\half\sqrt{s}$, where
$(E_B,\pvec_B)$ is the $B$ meson 4-momentum vector, and
all values are expressed in the \UfourS\ rest frame.
The resolution is $3.0\ \mevcc$ for \mes\ and in the range (12--32)~\mev
for \DE, depending on the decay mode.  We
require $5.25 <\mes<5.29 \gevcc$ and $|\DE|<0.2$ GeV.

Backgrounds arise primarily from 
continuum $\epem\ra\qqbar$ events ($q=u,d,s,c$).  We reduce these with
a requirement on the angle
\thetaT\ between the thrust axis of the $B$ candidate in the \UfourS\
rest frame and that of the rest of the charged tracks and neutral calorimeter
clusters in the event~\cite{thrust}.
The distribution is sharply
peaked near $|\costhr|=1$ for \qqbar\ jet pairs
and is nearly uniform for $B$ meson decays.  The requirement 
is $|\costhr|<0.9$.
For the \rhoz\ decays we also use $|\cos{\theta_{\rho}}|$ 
where the helicity angle $\theta_{\rho}$ is
defined   as the
angle between the momenta of a daughter pion and the
\etapr, measured in the \rhoz\  meson  rest frame. 
For \etagg\ decays we use $|\cos{\theta_{\eta}}|$ 
where the decay angle $\theta_{\eta}$ is
defined     as the
angle between the momenta of the most energetic daughter photon and
the 
\Bz\ meson, measured in the $\eta$  meson  rest frame. 
We require $|\cos{\theta_{\rho(\eta)}}| < 0.9$.
Events are retained only if they contain at least one charged track
in the decay products of the other $B$ meson ($B_{\rm tag}$)
from the $\Upsilon (4S)$ decay. This requirement improves the
precision of the determination of $B_{\rm tag}$ thrust axis.
The $\Bz\to\piz\KS\KS$ decay has background from $\Bz\to\overline{D^0}\KS$,
with $\overline{D^0}\to\piz\KS$, which has the same final state as the signal
mode. In order to suppress this background, we define $m(\piz\KS)$ as the
closer of the two invariant mass combinations to the nominal $D^0$
mass~\cite{PDG2008}.  By requiring $m(\piz\KS)$ to be outside the range
1.815--1.899~\gevcc, we veto 80\% of this background.

We obtain the signal event yields from unbinned extended maximum likelihood
(ML) fits.  The observables used in the fit are \DE, \mes, and  a Fisher
discriminant \xf.
The Fisher discriminant \xf~\cite{Fisher}  is a linear combination of
four event 
shape variables and $|\calT|$,  the absolute value of the continuous
output of a flavor tagging algorithm ~\cite{babarsin2betaprd}.
The event shape variables used for \xf\ are: the angles, with
respect to the  
beam axis, of the $B$ momentum and the $B$ thrust axis 
in the \UfourS\ frame, and the zeroth and second angular moments, $L_{0,2}$, 
of the energy flow about the $B$ thrust axis~\cite{Fisher1}.  The
moments are defined by 
$ L_j = \sum_i p_i\times\left|\cos\theta_i\right|^j$,
where $\theta_i$ is the angle, with respect to the $B$ thrust axis, of 
track or neutral cluster $i$, and $p_i$ is its momentum. The sum
excludes the $B$ candidate daughters.
We use a neural network based technique~\cite{babarsin2betaprd} to
 determine the flavor at decay of the $B_{\rm tag}$. 

The coefficients of \xf\ are chosen  to maximize the separation between
the signal and the continuum background. They are determined from
studies of Monte Carlo (MC)~\cite{geant} simulated signal data
and  off-peak data. 
Signal MC events are distributed uniformly across the Dalitz plot.
Correlations among the ML input observables are below 10\%.
The average number of candidates found per 
selected event is between $1.13$ and $1.22$, depending on the final
state.  We choose the candidate with the highest $B$ vertex  $\chi^2$
probability, determined from a vertex fit that includes both charged
and neutral particles~\cite{TreeFitter}. From 
 simulated events we find that 
this algorithm selects the correct candidate in 
(92--98)\% of the events containing multiple candidates, depending on
the final state,
and introduces negligible bias.

We use a MC simulation to estimate
backgrounds from other 
$B$ decays, including 
final states with and without charm. These contributions are negligible for 
the \etapepp\ mode. In all the other modes we introduce a non-peaking
\BB\ component in the fit. In the \fpizksks\ analysis we also
introduce a 
\BB\ background component that peaks in 
\mes\ and \DE, to take into account the main contribution to background
from $\Bz \to \KS \KS\KS$ decay mode.
We consider three components in the likelihood fit: signal, 
continuum, and \BB\ background. 
We have studied the possibility of misreconstruction of our $B$ candidates.
We divide signal events into two sub-components: correctly reconstructed
(COR) signal and self cross-feed 
(SCF) signal,  where at least one 
$B$ candidate daughter  has been exchanged with a particle from the
rest of the event. 
The signal component is split according to this classification.
The fractions of SCF events are fixed in the fit to
the values found in MC simulated events, which are in the range
(10--21)\%, depending on the final state.
For the \fpizksks\ decay mode, which has the lowest 
SCF fraction (6.6\%), we
use one signal component, comprising COR and SCF events.

For each event $i$ and  component $j$, we define the  probability
density function (PDF) 
\begin{equation}
{\cal P}^i_{j} = {\cal P}_j (\mes^i) {\cal  P}_j (\DE^i)  { \cal
  P}_j(\xf^i) \\ 
\end{equation}
and the likelihood function:
\begin{equation}
\Lik =  e^{-\left(\sum n_j\right)} \prod_{i=1}^N \left[\sum_j n_j  {\cal P}^i_{j} \right],
\end{equation}
where  $N$ is the number of reconstructed events and  $n_j$ is the number
of events in  
component $j$ which is returned by the fit.
We determine the PDF parameters from MC simulation
 of the  signal and \BB\ backgrounds, while we use 
  \mes\  and \DE\ sideband data ($5.25 < \mes\
<5.27$ \gevcc, $0.1<|\DE |<0.2$ \gev ) to model the PDFs of
continuum  background.

\begin{table*}[t]
\caption{
Fitted signal yield in events and fit bias in events (ev), detection
efficiency  \eff, daughter branching fraction product $\prod\calB_i$,
significance \signff\  and measured branching
fraction \calB\ with statistical error for each decay mode. For the
combined measurements (in bold) we give \signff\
(with systematic uncertainties included) and the  branching fraction
with statistical and systematic uncertainties with the  90\%
CL upper limit in parentheses.
}
\label{tab:results}
\begin{tabular}{lccccccc}
\dbline
Mode& \quad Yield (ev) \quad&\quad Fit bias (ev) &\quad \eff \quad &\quad
$\prod\calB_i$ (\%) \quad&\quad \signf \quad &\quad \bfemseven \quad  \\
\tbline
\bma{\fpizksks\ } &   $11.7\,^{+16.2}_{-14.5}$&$+1.0\pm 0.7$
&$17.5 $&$47.9 $&\bma{0.7}& \bma{2.7\,^{+4.2}_{-3.7}\pm 0.6
  \quad(<9)  } \\ 
\hline
~~\fetaggksks\  &   $3.2\,^{+9.0}_{-7.2}$ &$+1.1\pm
0.7$&$17.5 $&$18.8 $&$0.3$&$1.4\,^{+5.9}_{-4.7}$\\ 
~~\fetapppksks\  &   $2.2\,^{+5.5}_{-3.6}$
&$+0.2\pm0.6$&$12.0 $&$10.9 $&$0.5$&$3.3\,^{+9.0}_{-5.9}$\\ 
\bma{\fetaksks\ }&  &  &  & &\bma{0.5 } & \bma{2.1\,^{+4.7}_{-3.8} \pm
  1.2 \quad(<10) }   \\ 
\tbline
~~\fetapreppksks\ &   $2.4\,^{+4.7}_{-3.4}$  &$+0.1\pm 0.4$&$12.6 $&$8.4
$&$0.6$&$4.6\,^{+9.5}_{-6.9}$ \\  
~~\fetaprrgksks\ &   $13.4\,^{+16.1}_{-14.1}$  &$+4.7\pm1.1$&$15.9 $&$14.1
$&$0.6$&$8.3\,^{+15.4}_{-13.5}$ \\ 
\bma{\fetaprksks\ } &  &  &  & &\bma{0.8} & \bma{5.7\,^{+8.0}_{-6.5} \pm
  3.4 \quad(<20)  } \\ 
\dbline
\end{tabular}
\vspace{-5mm}
\end{table*}

We parameterize ${\cal  P}(\mes)$ as a Crystal Ball 
function~\cite{Crystal} 
for the COR and SCF signal sub-components, an ARGUS function
\cite{Argus} for 
continuum and non-peaking \BB\ background components, and by an 
ARGUS function plus an asymmetric Gaussian distribution
for peaking \BB\ background. 
The ${\cal  P}(\DE)$ distribution is described  by
an asymmetric Gaussian distribution plus an exponential tail (AGT)
\cite{Ignoto} for the COR signal sub-component, an asymmetric Gaussian
distribution plus a linear Chebyshev polynomial or an AGT for the SCF,
and Chebyshev polynomials for continuum and \BB\ background components. 
The distribution of \xf\  is described with an asymmetric Gaussian
distribution plus a Gaussian distribution for 
the COR signal sub-component, an  AGT function for SCF signal events,
an asymmetric Gaussian distribution
plus a linear Chebyshev polynomial for continuum, and an asymmetric
Gaussian distribution for \BB\ background sub-components. 

We allow  the continuum-background PDF
parameters  to float in the fit.
Large control samples of  \mbox{$\Bm\ra D^0(\KS\pip\pim\piz)\pim$} decays 
are used to verify the simulated  \DE\ and \mes\ resolution. 
Any bias in the fit, which mainly arises from neglecting the
correlations among the discriminating variables 
used in the likelihood function definition, is determined from a 
large set of simulated experiments.
For each experiment, the \qqbar\ background and
non-peaking \BB\ background are drawn
from the PDFs, and we embed  
the expected number of peaking \BB\  background  and signal events 
taken randomly from fully  simulated MC samples.

In Table~\ref{tab:results} we show, for each decay mode, the fitted
signal yields and their fit biases in numbers of events, the detection 
efficiencies, the product of daughter
branching fractions, 
the significance \signff, and the measured branching fractions.
  The detection efficiency is determined as the ratio of selected events in
simulation to the number generated. 
The significance is
given in units of $\sigma$.
We determine the corrected signal yields from the
fitted signal yields and their fit biases, estimated using
simulations.
We use these values, detection efficiencies, 
daughter branching fractions, and number of produced $B$ mesons,
assuming equal production  
rates of charged and neutral $B$ meson pairs, to 
compute the branching fractions.
The statistical error on the signal yield is  the change in 
the central value when the quantity $-2\ln{\Lik}$ increases by one 
unit from its minimum value. The significance is  the square root 
of the difference between the value of $-2\ln{\Lik}$ (with systematic 
uncertainties included) for zero corrected signal yield and the value
at its minimum. 
We combine results from different sub-decay modes by adding the values
of $-2\ln{\Lik}$.
In order to account properly for systematic uncertainties when combining
results from different sub-decays, we convolve the $\Lik$ of
each sub-decay mode with a Gaussian distribution with mean equal to
zero and width equal to the uncorrelated systematic uncertainty of
that decay mode.  
For the combined measurements we report the branching fractions,
the statistical significances and the 90\% confidence level (CL) 
upper limits. 
The 90\%  CL upper limit is taken to be the branching 
fraction below which lies 90\% of the total  likelihood integral 
in the positive branching fraction region.

\begin{figure}[!h]
\vspace*{0.5cm}
\hspace*{-0.5cm}
 \includegraphics[angle=0,scale=0.45]{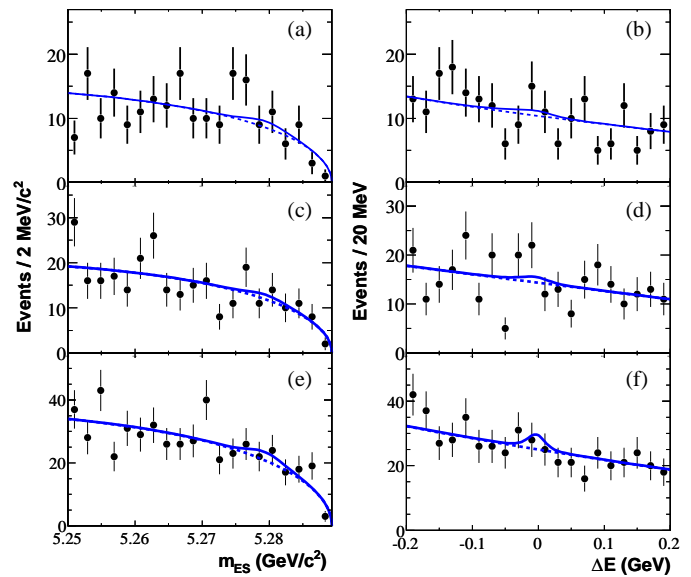} \\
\vspace*{-0.5cm}
\caption{
\Bz\  candidate \mes\ and \DE\ projections for 
\fpizksks (a,b), \fetaksks (c,d), and for \fetaprksks\ (e,f) 
with the sub-decay modes combined. Points with errors represent the
data, solid curves the full fit  functions and dashed curves the
background functions. These plots are  made with a requirement on the
likelihood in order to enhance signal to background ratio.
}
\label{f:projections}
\end{figure}

Figure~\ref{f:projections} shows projections of 
\fpizksks, \fetaksks, and \fetaprksks\   candidates 
onto \mes\ and \DE\ for the  subset of candidates for which the signal
likelihood 
(computed without the variable plotted) exceeds a mode-dependent
threshold.

The main sources of systematic error include uncertainties in the
detection efficiencies, the
 PDF  parameters,  and the maximum likelihood fit bias. We assign 
 systematic uncertainties (13--20\%) on the detection efficiencies due
 to non-uniformity of the efficiencies over the Dalitz plot. 
This contribution is taken to be the ratio between the standard
deviation of the efficiency distribution over the Dalitz Plot to its
mean value. 
For the signal, the uncertainties in the  PDF parameters are estimated
by comparing MC and data control samples.  Varying the 
signal PDF parameters within these uncertainties, we estimate the yield
uncertainties  of \mbox{0--2} events, depending on the mode.  The uncertainty
from the fit bias is taken as the sum in quadrature of one-half the correction
(1--3 events) plus the statistical uncertainty on the
correction itself.
We assign a systematic error of \mbox{0.1--0.4} events, depending on
the mode, due to non-uniformity of the SCF fraction over the Dalitz plot.
Uncertainties of the efficiency found from auxiliary studies include
$0.8\%\times\;N_t$ where $N_t$ is the number of tracks 
in the $B$ candidate. A systematic uncertainty of $1.8\%$ and $3.0\%$
is assigned to the single photon and \piz/$\etagg$ meson  reconstruction
efficiencies, respectively.
There is a  systematic error of 0.9\% for the
reconstruction efficiency of each \KS.
  The uncertainty on the total number of \BB\ pairs in the
data sample is 1.1\%.  Uncertainties  on the 
$B$ daughter branching-fraction products (3.5--4.9)\%  are taken from 
Ref.~\cite{PDG2008}.

In conclusion
we have searched for the \Bz\  decay modes to \fpizksks, \fetaksks\ and
\fetaprksks\ with a sample of $467 \times 10^{6}$ \BB\ pairs.
We find no significant signals and set  90\% CL  upper 
limits for the branching fractions: 
$\calB(\Bz \to \fpizksks\ ) <9 \times 10^{-7}$, $\calB (\Bz
\to\fetaksks) <10  \times 10^{-7}$, 
and $\calB (\Bz\to\fetaprksks\ ) <20  \times 10^{-7}$.

We are grateful for the excellent luminosity and machine conditions
provided by our \pep2\ colleagues, 
and for the substantial dedicated effort from
the computing organizations that support \babar.
The collaborating institutions wish to thank 
SLAC for its support and kind hospitality. 
This work is supported by
DOE
and NSF (USA),
NSERC (Canada),
CEA and
CNRS-IN2P3
(France),
BMBF and DFG
(Germany),
INFN (Italy),
FOM (The Netherlands),
NFR (Norway),
MES (Russia),
MEC (Spain), and
STFC (United Kingdom). 
Individuals have received support from the
Marie Curie EIF (European Union) and
the A.~P.~Sloan Foundation.

\renewcommand{\baselinestretch}{1}


\begin{thebibliography}{99}
\bibitem{Ignoto}
We use the function $f(x)=N \exp \left(
\frac{-(x-\mu)^2}{2\sigma^2_{L,R} + \alpha_{L,R} (x-\mu)^2} \right)$,
where $\mu$ is the peak position of the distribution, $\sigma_{L,R}$
are the left and right widths, $\alpha_{L,R}$ are the left and
right tail parameters, and $N$ a normalization factor.
\end{thebibliography}
\end{document}